\documentclass[a4paper,11pt]{article}
\usepackage{pos}

\def\beq{\begin{equation}}
\def\eeq{\end{equation}}
\def\bea{\begin{eqnarray}}
\def\eea{\end{eqnarray}}
\def\qon#1{q_{#1,0}^{(+)}}
\def\Eq#1{Eq.~(\ref{#1})}
\def\qb{\mathbf{q}}

\def\ii{\imath 0}

\title{Understanding IR singularities in the Loop-Tree Duality}

\author*[a]{Germ\'an Rodrigo}
\author[a]{Leandro Cieri}
\author[b]{Prasanna K. Dhani}
\author[c]{Roger J. Hern\'andez-Pinto}
\author[a]{Jorge J. Mart\'{\i}nez de Lejarza}
\author[a,c]{Salvador A. Ochoa-Oregon}
\author[a]{Konstantinos Pyretzidis}
\author[c]{Selomit Ram\'{\i}rez-Uribe}
\author[a]{David F. Renter\'{\i}a-Estrada}
\author[d]{Andr\'es E. Renter\'{\i}a-Olivo}
\author[a]{Jos\'e R\'{\i}os-S\'anchez}
\author[a,e]{German F.R. Sborlini} 
\author[a]{Surabhi Tiwari}
\author[c]{Juan P. Uribe-Ram\'{\i}rez}

\affiliation[a]{Instituto de F\'{\i}sica Corpuscular, Universitat de Val\`{e}ncia -- Consejo Superior de Investigaciones Cient\'{\i}ficas, Parc Cient\'{\i}fic, E-46980 Paterna, Valencia, Spain}

\affiliation[b]{Physik Institut, Universit\"at Z\"urich, Winterthurerstrasse 190, CH-8057 Z\"urich, Switzerland}

\affiliation[c]{Facultad de Ciencias Naturales y Exactas, Universidad Aut\'onoma de Sinaloa, Ciudad Universitaria, CP 80000 Culiac\'an, Mexico}

\affiliation[d]{Donostia International Physics Center (DIPC), 20018 Donostia-San Sebasti\'an, Basque Country, Spain}

\affiliation[e]{Departamento de F\'isica Fundamental e IUFFyM, Universidad de Salamanca, 37008 Salamanca, Spain}

\emailAdd{german.rodrigo@csic.es}

\abstract{One of the main advantages of the Loop-Tree Duality representation of scattering amplitudes is that it makes the origin of infrared and threshold singularities particularly transparent. This talk reviews recent progress in describing how singularities emerge and cancel at the level of scattering and vacuum amplitudes, discusses a novel strategy to efficiently construct finite integrals, and presents a complementary perspective based on encoding the underlying causal and singular structure in terms of qubits and quantum circuits.
}

\FullConference{Loops and Legs in Quantum Field Theory (LL2026)\\
12-17, April, 2026\\
Bayreuth, Germany\\}

\begin{document}
\maketitle

\section{Introduction}

In the current landscape of collider physics, high-precision theoretical predictions in Quantum Field Theory (QFT) at high perturbative orders are essential for meaningful comparisons with experimental data. Perturbative calculations in QFT rely on multiloop Feynman integrals, which typically possess a highly intricate mathematical structure further complicated by the presence of singularities. These include ultraviolet (UV) singularities in the high-energy regime, soft and collinear singularities in the corresponding infrared (IR) limits, and threshold singularities arising when internal particles simultaneously go on-shell, thereby describing a physical scattering process.

Beyond their practical applications in collider physics, multiloop Feynman integrals are also of intrinsic interest as mathematical objects. As we push toward higher loop orders, their singularity structure becomes increasingly involved and automated public codes, such as \texttt{pySecDec}~\cite{Binoth:2000ps,Borowka:2017idc} and \texttt{Fiesta5}~\cite{Smirnov:2021rhf}, struggle with substantial computational bottlenecks. These bottlenecks manifest themselves in numerical instabilities and in the increasing difficulty of reducing higher-order integrals to a basis of master integrals.

Focusing on finite integrals, i.e. Feynman integrals free of UV, IR, and threshold singularities, may initially be seen as a way to sidestep the main challenges of multiloop calculations. Nevertheless, finiteness does not imply simplicity: such integrals can exhibit intricate kinematic dependence and highly non-trivial analytic structures. Their study therefore provides a controlled setting for isolating the intrinsic difficulties of loop integration and for benchmarking analytic and numerical methods before extending them to singular cases. In particular, finite integral bases may facilitate the optimal construction of master integrals and alleviate the computational bottlenecks associated with large-scale IR subtractions by absorbing IR poles into coefficients, thereby smoothing the $d\to 4$ limit.

Therefore, the identification and construction of bases of finite Feynman integrals at high perturbative orders, rather than a theoretical curiosity, has been advocated as a promising strategy for achieving numerically stable computations. Established strategies to construct finite Feynman integrals include dimensional shift~\cite{Bern:1992nf}, working in the Euclidean region~\cite{Panzer:2014gra, vonManteuffel:2014qoa} or introducing raised propagators, also called “dots”~\cite{vonManteuffel:2012np,Studerus:2009ye}. Another efficient approach involves using suitable numerators to cancel IR singularities directly at the integrand level~\cite{Agarwal:2020dye,Gambuti:2023eqh,delaCruz:2024xsm,Figueiredo:2026ksz}. Further developments comprise an initial set of finite integrals to iteratively construct an Integration-By-Parts~(IBP) basis in which singularities are absent in the reduction coefficients~\cite{DeAngelis:2025agn}. Also, Hodge theory provides a unified framework for treating the geometries associated with finite Feynman integrals~\cite{Weinzierl:2026gik}.

In this talk, we review a novel strategy that leverages the ability of the Loop–Tree Duality~(LTD) to provide a transparent framework for understanding the origin of infrared and threshold singularities directly at the integrand level. We discuss the three distinct routes to finiteness presented in Ref.~\cite{Dhani:2026cxx}, with particular emphasis on the third, which efficiently circumvents the UV scaling induced by the introduction of numerators. We further examine how to select acyclic configurations, which provide the foundation for understanding the singularity structure of Feynman integrals within LTD, in a Quantum Computing approach. 

\section{Feynman versus LTD Representations}

In the Feynman framework, loop integrations are performed over the full $d$-dimensional Minkowski spacetime. LTD reformulates these integrations by applying the Cauchy's residue theorem to integrate out one component of each loop four-momentum~\cite{Catani:2008xa,Aguilera-Verdugo:2020set}. When this component is the temporal energy component~\cite{Buchta:2015wna,Runkel:2019yrs,Capatti:2019ypt,Soper:1998ye}, the integration domain is projected onto the ($d-1$)-dimensional Euclidean space of the loop three-momenta, yielding an LTD representation that is also manifestly causal~\cite{Aguilera-Verdugo:2020kzc,Ramirez-Uribe:2020hes}.

In this LTD causal representation, the standard Feynman propagators are replaced by causal propagators, which are expressed as linear combinations of external energies, $k_{(1,n),0}$, and internal on-shell energies
\beq
\frac{1}{\lambda^\pm_{i_1\cdots i_n}} = \frac{1}{\sum \qon{i_s} \pm k_{(1,n),0}}~, \label{eq:causal}
\eeq
where the on-shell energies are defined as $\qon{i_s}=\sqrt{\qb_{i_s}^2+m_{i_s}^2-\ii}$. The complex $\ii$-prescription is inherited from the branch structure of the original Feynman propagator. The causal propagators significantly simplify the singularity structure of the integrand. Divergences arise when their denominators vanish, with \Eq{eq:causal}, which characterizes only configurations associated with physical thresholds or collinear and soft limits. In the LTD representation, unphysical or nonpinched singularities of the integrand are absent. This formulation avoids the often opaque manifestation of singularities in the Feynman representation and enables the precise identification of individual singular configurations.

The LTD representation is also central in interpreting Feynman propagators as qubits~\cite{Ramirez-Uribe:2021ubp,Clemente:2022nll,Ramirez-Uribe:2024wua,Ochoa-Oregon:2025opz}. In the covariant framework, the Feynman propagator $G_{\rm F}(q_i)$ describes a quantum superposition of propagation in both directions between two interaction vertices. This property can be expressed formally as
\beq
G_{\rm F}(q_i)= \frac{1}{q_i^2 - m_i^2 + \imath 0}\equiv \frac{1}{2}\left(|0\rangle+|1\rangle \right), 
\eeq
where $|0\rangle$ and $|1\rangle$ denote the state of a particle moving forward or backward in time. LTD effectively measures this state through the application of Cauchy’s residue theorem, resolving the superposition into directed, on-shell states in which the directions of propagation are fixed. By reformulating the problem in the Euclidean space of the loop three-momenta, LTD provides a manifest exploration of causality, as the resulting representation is constructed entirely from acyclic configurations.

\section{Causality, Acyclicity, and Graph Theory}

The synergy between causality and acyclicity in graph theory constitutes a cornerstone of multiloop LTD calculations, providing an efficient framework for identifying the singularity structure of Feynman integrals directly at the integrand level. By restricting the analysis to acyclic configurations, one excludes incompatible causal assignments and ensures that the remaining singular configurations correspond to physically meaningful thresholds. This graph theory-based perspective facilitates the systematic classification of singularities. If a particle’s momentum flow in a loop returns to its emission point, the configuration is cyclic, representing a nonphysical breaking of causality. Such cyclic configurations are the source of spurious singularities in the covariant Feynman representation that requires the summation of both cyclic and acyclic terms.

\begin{figure}[t]
\centering
\includegraphics[width=.75\linewidth]{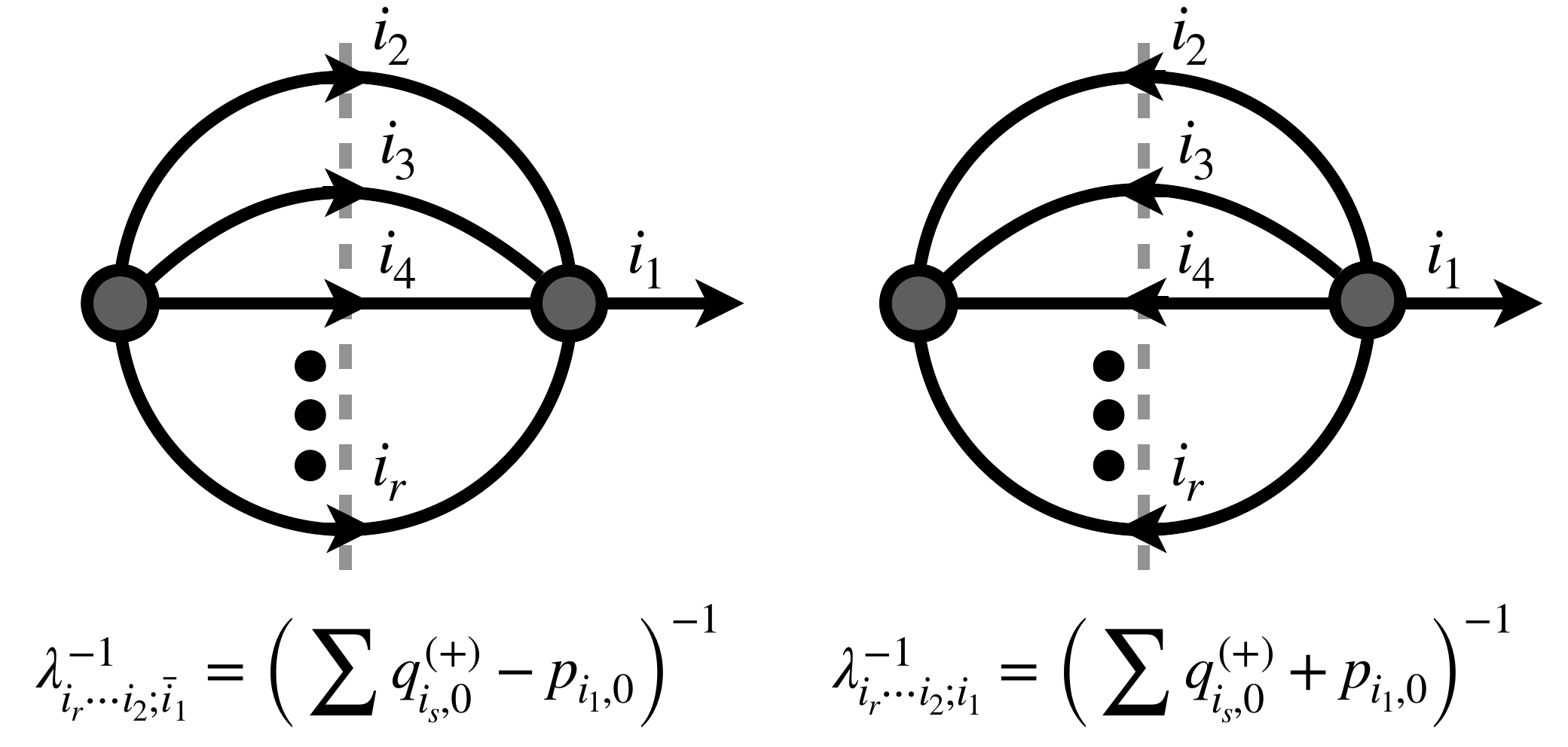}
\caption{Causal propagators representing configurations in which the momentum flows of the internal particles are aligned (left) or counter-aligned (right) with the external momentum. The line intersecting the internal particles divides the amplitude into two subamplitudes. The gray blobs represent any multiloop subgraphs. The right configuration is finite, assuming $p_{i_1,0}>0$.}
\label{fig:causalpropagators}
\end{figure}

\begin{figure}[t]
\centering
\includegraphics[width=.3\linewidth]{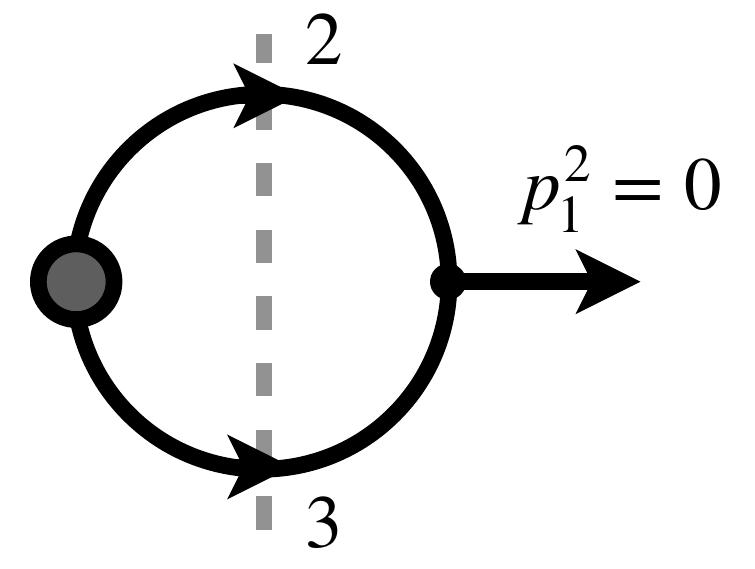}
\includegraphics[width=.3\linewidth]{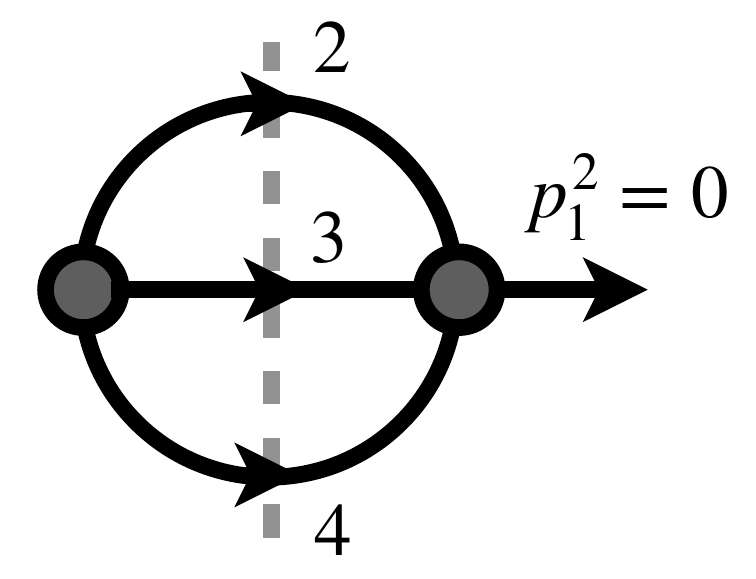}
\includegraphics[width=.3\linewidth]{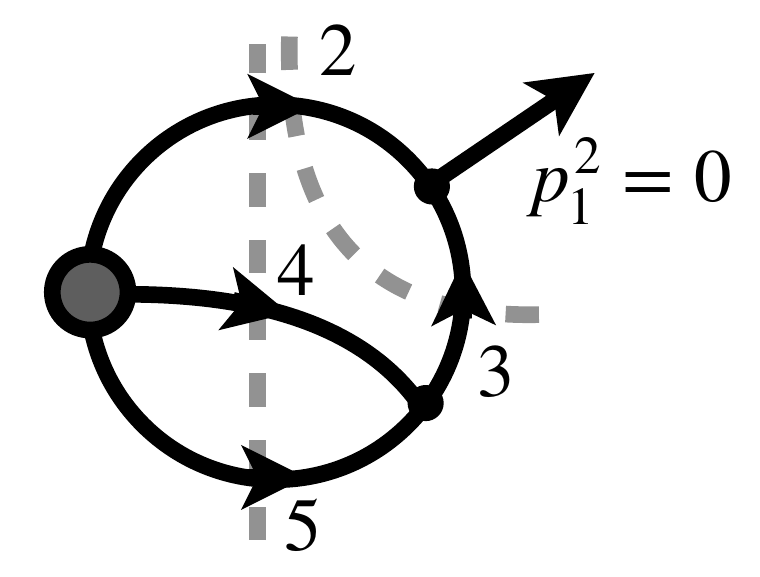}
\caption{Double collinear configuration $\lambda_{32;\bar 1}^{-1} = {\cal O}(k_\perp^{-2})$, triple collinear configuration $\lambda_{432;\bar 1}^{-1} = {\cal O}(k_\perp^{-2})$, and overlapping triple and double collinear configuration $\lambda_{32;\bar 1}^{-1} \lambda_{542;\bar 1}^{-1} = {\cal O}(k_\perp^{-4})$. The internal particles are massless and their momentum flows are aligned with the external momentum. For massive particles, these configurations lead to physical thresholds.}
\label{fig:collinear}
\end{figure}

Figure~\ref{fig:causalpropagators} provides a graphical interpretation of the causal propagator in \Eq{eq:causal}. The momentum flows of the internal particles involved are always aligned or counter-aligned with the momentum of the external particle. This leads to two complementary contributions: $\lambda^{-1}_{i_r\cdots i_2; i_1} = \left(\sum \qon{i_s} + p_{i_1,0}\right)^{-1}$ is finite, assuming $p_{i_1,0}>0$, while $\lambda^{-1}_{i_r\cdots i_2; \bar i_1} = \left(\sum \qon{i_s} - p_{i_1,0}\right)^{-1}$ may develop a singularity for $\lambda_{i_r\cdots i_2; \bar i_1} \to 0$. Notice that if a subset of the internal particles propagates in the opposite direction a cyclic configuration would be generated, which would be encoded by flipping the sign of the corresponding on-shell energies. This would also generate a singularity which, however, is spurious and is absent in the LTD representation. For massive particles, the causal propagator $\lambda^{-1}_{i_r\cdots i_2; \bar i_1}$ represents in general a physical threshold, although the magnitude of the masses and external energy determine if kinematics actually allows $\lambda_{i_r\cdots i_2; \bar i_1}$ to vanish. In the Euclidean space of the loop three-momenta, threshold singularities occur in an ellipsoid surface. At the two-loop level, thresholds involving two loop momenta form a compact five-dimensional surface.

\begin{figure}[t]
\centering
\includegraphics[width=.78\linewidth]{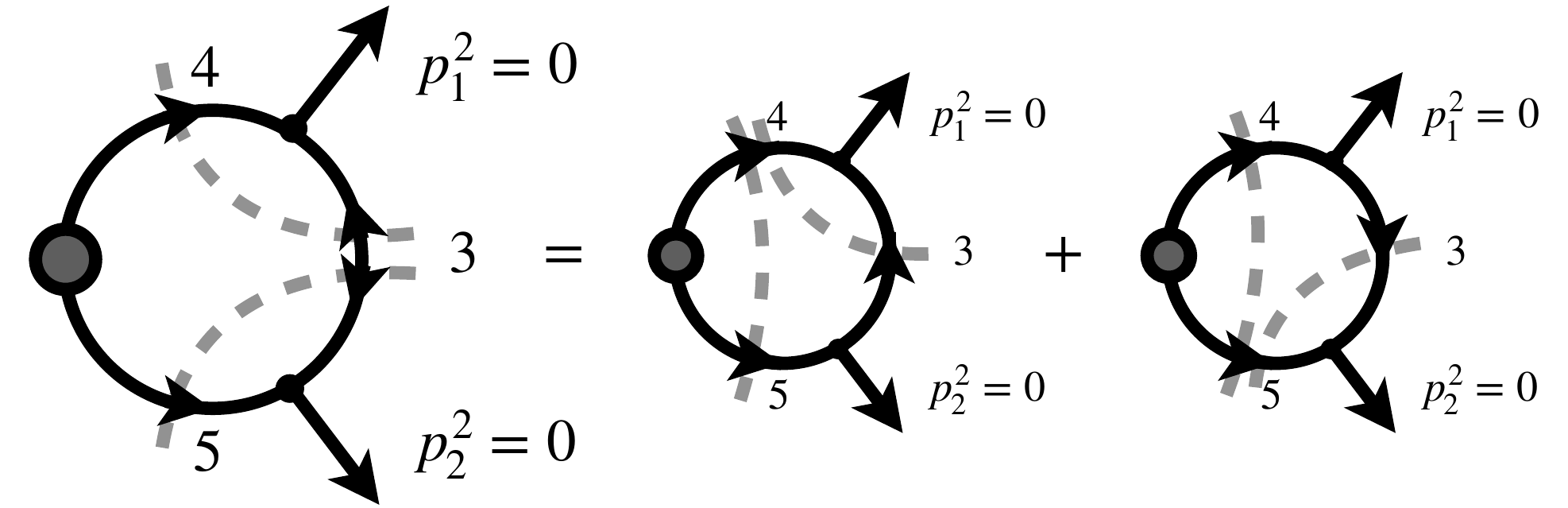}
\caption{Soft singularity: $\left(\lambda_{45;\overline{12}} \left( \lambda_{43;\overline 1} + \lambda_{53;\overline 2}\right)\right) ^{-1}$. The two collinear configurations on the left become compatible in the soft limit ($\lambda_3 \to 0$).}
\label{fig:soft}
\end{figure}

For massless particles, $\lambda^{-1}_{i_r\cdots i_2; \bar i_1}$ may represent a soft or collinear configuration, again kinematics dependent. Explicit examples of double collinear and triple collinear configurations are shown in Fig.~\ref{fig:collinear}. We can also identify configurations exhibiting overlapping singularities, which are represented by products of causal propagators. Soft singularities (Fig.~\ref{fig:soft}) require the contribution of two, in principle, causal incompatible collinear configurations that however become compatible in the soft limit. In the Euclidean space of the loop three-momenta, collinear singularities occur in a one-dimensional finite segment, which is the result of collapsing an ellipsoid surface to one dimension. 

In the absence of external particles, the only possible causal propagators are of the form $\lambda^{-1}_{i_r\cdots i_1} = \left(\sum \qon{i_s}\right)^{-1}$, which cannot become singular unless all on-shell energies vanish. As for regular scattering amplitudes, a line intersecting the internal particles involved in the causal propagator divides the amplitude into two subamplitudes. These two properties have motivated the recent proposal~\cite{Ramirez-Uribe:2024rjg} of using vacuum amplitudes in LTD as the optimal building blocks to assemble theoretical predictions at high-energy colliders. Explicit examples, such as the decay rate of the Higgs boson to heavy quarks at second order in perturbation theory, have been presented in Ref.~\cite{LTD:2024yrb}, together with an illustrative example at the next order. The good behaviour of the integrand has also enabled us to implement and integrate the integrand expression obtained in a Quantum Computing approach~\cite{deLejarza:2024scm,Pyretzidis:2025stx}.

Specific final states are generated as residues in the causal propagators, the so-called phase-space residues,
\beq
d\sigma_{{\rm N}^k{\rm LO}}
\sim \int_{\vec{\ell}_1 \cdots \vec{\ell}_{\Lambda-2}} \sum {\rm Res} \left( {\cal A}_{\rm D}^{(\Lambda, {\rm R})}, \lambda_{i_r \cdots i_2 i_1 a b} \right) \delta\left( \lambda_{i_r \dots i_2 i_1} - p_{a,0}- p_{b,0}\right)~,
\eeq
in such a way that loop and tree-level contributions are evaluated under the same integral. Central to this approach is the analytic continuation to negative values of the on-shell energies of particles identified as incoming, which also ensures the correct momentum flow of the phase-space residues interpreted as the interference of two amplitudes with external particles. Since the vacuum amplitude is free of IR and threshold singularities, and all contributions are generated coherently from the same vacuum amplitude, which acts as a kernel, the sum over all phase-space residues exhibits a local cancellation of IR singularities. UV singularities are locally cancelled by appropriate UV counter-terms. Threshold singularities may also match between phase-space residues.

\section{Finite Feynman integrals} 
\label{sec:finite}

The first two strategies presented in Ref~\cite{Dhani:2026cxx} for constructing finite integrands involve cancelling IR singularities directly at the integrand level using an appropriate numerator or by manipulating the propagator powers and spacetime dimensionality. We define a numerator Ansatz either in the Feynman representation, as a polynomial of scalar products of loop and external momenta with arbitrary coefficients, or as a polynomial of on-shell energies, also with arbitrary coefficients. Then, we transform the target integrand into its LTD representation, and impose that all residues on singular causal propagators vanish: 
\beq
{\rm Res} \left( {\cal A}_{\rm D}^{(\Lambda)}, \lambda_{i_r\cdots i_2;\bar i_1}\right) = 0~. 
\eeq
The result is a set of polynomials in the on-shell energies with undetermined coefficients, which yields a system of linear equations for the coefficients of the corresponding monomials. Solving this system determines the coefficients of the numerator that render the integral finite.

Feynman propagators raised to a power $a$ lead to raised causal propagators, such that $\left( \lambda_{i_r\cdots i_2;\bar i_1}\right)^{-a} = {\cal O} (k_\perp^{-2a})$ in the collinear limit. In that case, we must impose that the residues of the higher-order poles vanish in $d+a$ spacetime dimensions:
\beq
\left. {\rm Res} \left( (\lambda_{i_r\cdots i_2;\bar i_1})^{a-1}{\cal A}_{\rm D}^{(\Lambda)}, \lambda_{i_r\cdots i_2;\bar i_1}\right) \right|_{d+a} = 0~. 
\eeq
While effective, these first two strategies, which are still rooted in the Feynman representation as the target integrand, frequently produce integrands with rapid UV behaviour.

The third and the most advanced strategy consists of an intrinsic LTD-based construction. This approach builds integrands directly from the LTD representation using only causal denominators that correspond to non-singular configurations. By counter-aligning the direction of propagation for internal particles with the external momenta, we ensure that the corresponding LTD representation is IR finite, threshold free, or both. Unlike the multiplicative numerator method, which relies on high-rank polynomials that worsen the UV scaling, the intrinsic LTD method naturally leads to integrands with significantly milder UV behavior. In addition, this method allows us to easily design threshold-free integrands, eliminating the need for complex numerical procedures such as contour deformation, $\ii$-extrapolation, or shifts to the Euclidean region, making them ideal for high-order topologies benchmarking.

\section{Benchmark finite integrals}

\begin{figure}[t]
\centering
\includegraphics[width=.75\linewidth]{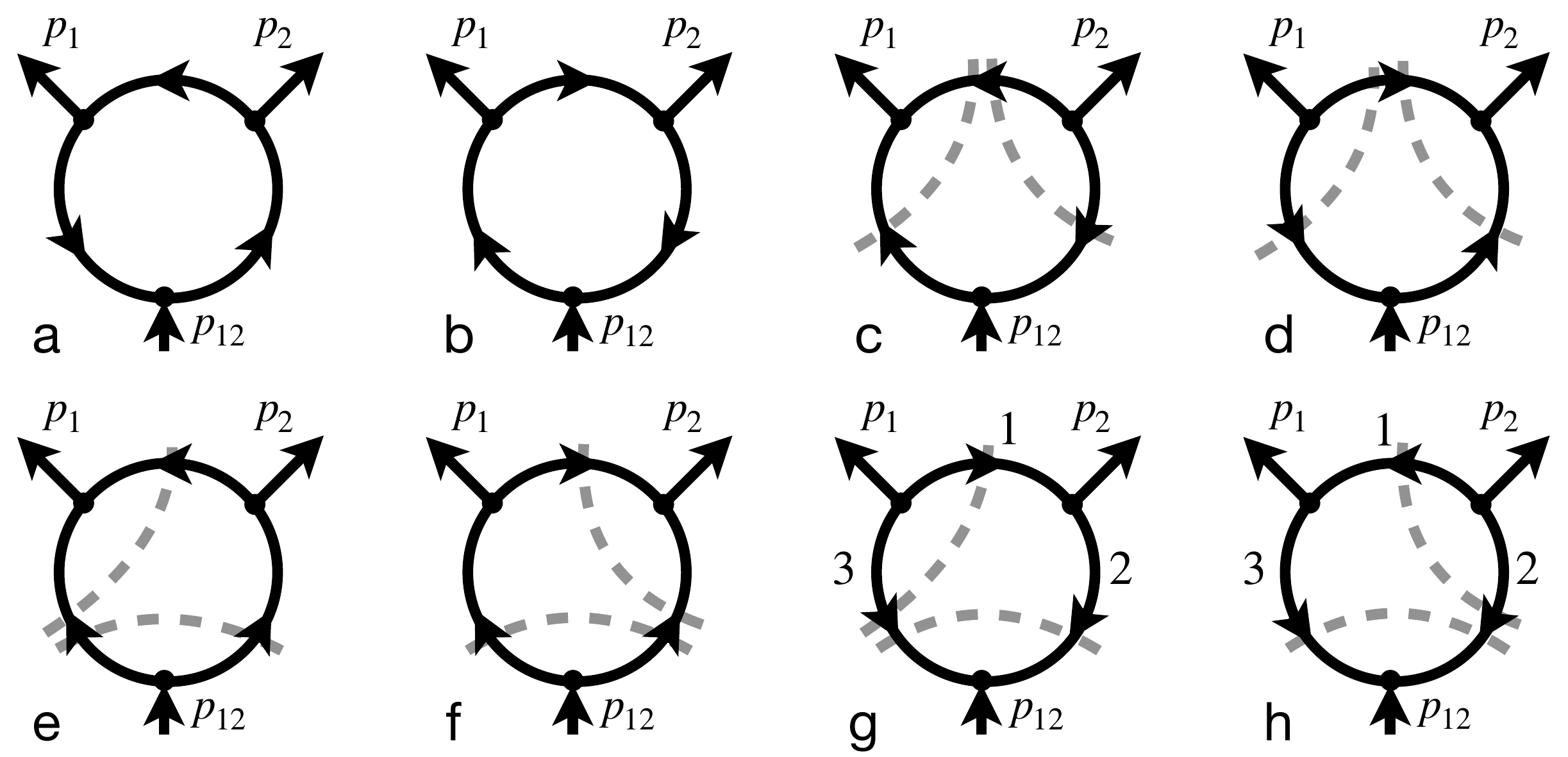}
\caption{The eight directed states of the one-loop three-point function: a) and b) are cyclic and therefore absent from the LTD representation; c) and d) are collinear; e) and f) are collinear, soft and threshold singular; g) and h) are finite.}
\label{fig:LTDtriangle}
\end{figure}

To illustrate the construction of finite integrals, we consider the one-loop three point function. In the Feynman representation 
\beq
{\cal A}^{(1)} = \int_{\ell_1} {\cal N}^{(1)}_{(1,1,1)} G_{\rm F} (q_1,q_2,q_3)~,
\label{eq:triangle}
\eeq
with $q_1 = \ell_1 + p_1$, $q_2 = \ell_1 + p_{12}$ and $q_3 = \ell_1$. There are three Feynman propagators and therefore eight directed states, as shown in Fig.~\ref{fig:LTDtriangle}. The first two are cyclic and therefore absent from the LTD representation, although they could generate unphysical nonpinched singularities in the Feynman representation. Our Ansatz numerator is 
\bea
{\cal N}^{(1)}_{(1,1,1)}(\ell_1) &=& \alpha_0 + \alpha_1 \, \ell_1\cdot p_1 + \alpha_2 \, \ell_1\cdot p_2 + \alpha_3 \, (\ell_1\cdot p_1)(\ell_1\cdot p_2) \nonumber \\ &+&  \alpha_4 \, (\ell_1\cdot p_1)^2 + \alpha_5 \, (\ell_1\cdot p_2)^2 + \alpha_6 \, \ell_1^2~.
\eea

As explained in Sec.~\ref{sec:finite}, we transform \Eq{eq:triangle} into its LTD representation, and nullify the following residues
\beq
{\rm Res}\left({{\cal A}_{{\rm D}(1,1,1)}^{(1)},\lambda_{31;\bar1}}\right) = 0~, \qquad
{\rm Res}\left({{\cal A}_{{\rm D}(1,1,1)}^{(1)},\lambda_{32;\bar2}}\right) = 0~.
\eeq
They suppress the configurations where the external momentum $p_1$ (respectively $p_2$) is collinear to the internal momentum $q_1$ (respectively $q_2$). By solving the corresponding system of linear equations, we obtain, with relabelled coefficients, 
\beq
{\cal N}^{(1, {\rm UV})|d}_{(1,1,1)}(\ell_1) = \alpha_0 \, (q_1\cdot p_1)(q_2\cdot p_2) + \alpha_1 \, q_1^2~.
\label{eq:numfinite}
\eeq
Notice, however, that although the numerator in \Eq{eq:numfinite} generates an IR finite integrand, it also scales in the UV, leading to a UV singularity in $d=4-2\epsilon$ spacetime dimensions. The only option to render the integral also UV finite is to raise some of the propagators, e.g. ${\cal N}^{(1) | d}_{(2,1,1)} = \alpha_0 \, \left(q_1^4-q_2^2 q_3^2\right)$, or ${\cal N}^{(1) | d+2}_{(1,1,2)} =  \alpha_0 \, (q_1\cdot p_1)$ in shifted spacetime dimensions. 

A better strategy is given by the intrinsic LTD-base construction, which in addition does not require any calculation. A graphical inspection of the directed configurations is sufficient. We just need to identify those configurations where the momentum flows of the internal particles are not aligned with the momentum of the external particles. For the one-loop three-point function in Fig.~\ref{fig:LTDtriangle}, only the last two configurations fulfil this condition, which are directly translated into the LTD expression:
\beq
{\cal A}^{(1)}_{{\rm D}(1,1,1)} = 
\frac{1}{x_{123}} \left(
\frac{r_1}{\lambda_{31;1}} +
\frac{r_1'}{\lambda_{12;2}}  \right) \frac{1}{\lambda_{23;12}}~, \qquad x_{123} = \prod_{s=1}^3 2q_{i_s,0}^{(+)}~,
\eeq
where $r_1$ and $r_1'$ are arbitrary polynomials of degree one in the on-shell energies. We refer to Ref.~\cite{Dhani:2026cxx} for more benchmark finite integrands at higher loop orders.

\section{A Quantum Computing approach} 

The task of identifying the subset of acyclic configurations and, in particular, those that yield finite integrals, within the complete set of possible momentum-flow configurations in multiloop Feynman diagrams can be recast as an unstructured search over an exponentially large configuration space. This type of exercise is well suited to a Quantum Computing approach. The construction is based, first, on the formal identification of Feynman propagators with qubits~\cite{Ramirez-Uribe:2021ubp, Clemente:2022nll}, as specified in \Eq{eq:causal}. Its second key ingredient is the use of the well-known multicontrolled Toffoli gate to probe cycles~\cite{Ramirez-Uribe:2024wua}. This gate acts on several control qubits and one target qubit; when all control qubits are in the $|1\rangle$ state, it flips the state of the target qubit. For Feynman diagrams, the control qubits represent the propagators forming a loop, and the target qubit is an ancillary qubit used in a quantum circuit to tag the cyclic configurations. 

The efficiency of a quantum search algorithm depends critically on the quantum resources required, i.e., the number of qubits and quantum gates. 
Given the close connection with graph theory, Ref.~\cite{Ochoa-Oregon:2025opz} has analysed the logical relations among cycles implemented by multi-controlled Toffoli gates and identified those that are mutually exclusive. These relations are represented through the adjacency graph of Mutually Exclusive Clauses (MECs), thereby recasting the quantum oracle optimisation into a Minimum Clique Partition problem: finding the smallest set of cliques that covers all vertices of the MEC graph. This procedure reduces significantly the number of ancillary qubits, as all MECs within a clique can be encoded using a single ancillary qubit.

\section{Conclusions and Outlook}

The LTD framework provides a streamlined and physically intuitive approach to classifying IR-finite multiloop Feynman integrands. By rendering the origin of singularities transparent, LTD avoids the computational complexity associated, for example, with the Landau analysis of Feynman integrals in the Feynman representation. The intrinsic LTD-based construction introduced in Ref.~\cite{Dhani:2026cxx} constitutes a significant advance, as it admits a graphical interpretation that requires no explicit calculations to construct finite integrals. In addition, it circumvents the UV scaling induced by numerator factors and can be formulated to yield threshold-free integrands directly. The key element of this construction is to enforce acyclic configurations where the momentum flows of the internal particles are counter-aligned with the momentum of the external particles. The selection of the acyclic configurations is also a task that is well suited for a Quantum Computing approach. 

As precision requirements in particle physics continue to increase, the ability to construct a basis of finite, threshold-free integrands directly within the LTD framework may play a central role in the efficient evaluation of multiloop scattering amplitudes.

\section*{Acknowledgments}
This work is supported by the Spanish Government and ERDF/EU - Agencia Estatal de Investigaci\'on MCIN/AEI/10.13039/501100011033,  Grants No. PID2023-146220NB-I00, No. EUR2025-164820, and No. CEX2023-001292-S. LC is supported by Generalitat Valenciana plan GenT program (CIDEGENT/2020/011). PKD is supported by the Swiss National Science Foundation (SNSF) under contracts 200020$\_$219367. RJHP thanks SECIHTI for the support received through Project CBF2023-2024-268 from the 2023-2024 frontier science call. KP is funded by AEI, Grant No. CEX2023-001292-S. DFRE is supported by Generalitat Valenciana, Grant No. CIGRIS/2022/145. SRU acknowledges support from SECIHTI through the Sistema Nacional de Investigadoras e Investigadores (SNII). JRS is funded by AEI, Grant No. PREP2023-001474. ST is supported by Generalitat Valenciana, Grant No. CIAPOS/2024/469. JPUR is funded by the Spanish Government and NextGenerationEU through Quantum Spain.

\bibliographystyle{JHEP}
\bibliography{2026_LL_rodrigo}

\end{document}